# A Digital Forensics Case Study of the DJI Mini 3 Pro and DJI RC

Aaron Taylor

*Abstract*— The consumer drone market is rapidly expanding with new drone models featuring unique variations of hardware and software. The rapid development of drone technology and variability in drone systems can make it difficult for digital forensic investigators and tools to keep pace and effectively extract and analyse digital evidence from drones. Furthermore, the growing popularity of drones and their increased use in illegal and harmful activities, such as smuggling, espionage, and even terrorism, has led to an increase in the number of drone forensic cases for authorities to manage. To assist forensic investigators, a static digital forensic case study was conducted on two drone devices recently released by Da-Jiang Innovations (DJI): the Mini 3 Pro drone, and its remote controller, the DJI RC. The study discovered the presence of several digital artefacts on both devices, including recorded media, flight logs, and other information that could help investigators trace the drone's usage and identify its operator. Additionally, this paper explored several methods for extracting and visualising the drone's flight history, and highlights some of the potential methods used to limit, obscure, or remove key types of digital evidence.

## I. INTRODUCTION

Drones are unmanned aircraft vehicles (UAVs) that typically operate autonomously by an onboard computer or controlled remotely by a human pilot [1]. Drones were typically limited to police, search and rescue, and military operations [2], [3]; however, in recent years they have become increasingly popular in commercial and recreational settings due to various factors, such as declining cost, changing public perception, and expanding range of applications ranging from recreational flying, agricultural spraying, parcel delivery, through to aerial photography and surveying [2], [4].

While legitimate use of drones has increased, so too has their involvement in illegal activity and accidents [5]. The Federal Aviation Administration (FAA) has reported an increase in drone accidents and incidents in the last two years [6]. Additionally, there are reports of drones being used for illegal activities, such as drug smuggling, gathering intelligence for criminal purposes, disrupting critical infrastructure, and even terrorism [3], [4], [7], and [8]. The increasing number of drone incidents and illegal use of drones has inevitably increased the number of Digital Forensic (DF) investigations involving drones [3], [9].

Drones present several challenges to the DF field. They often feature a complex array of technologies, including wireless communication, GPS, and cameras, which can make it challenging for investigators conducting a DF investigation [8], [10]. There is also a lack of modern DF tools specifically developed for drones [4], [10], and this, combined with the fast-paced development of drone technologies, can rapidly render available tools obsolete and unfit for purpose [4], [5]. Additionally, Drone Forensics (DRF) is a relatively new discipline within the DF field. Only recently, Interpol published guidelines for DRF investigators [8], and researchers [4], [10] have proposed novel methodologies for the DRF field. Despite these efforts, there still remains a lack of universally agreed methodologies and approaches [9], [11], presenting further challenges for DF investigators to navigate.

To ensure DF investigators are effective at conducting DRF investigations, they require specialist knowledge and skills [3], and should be aware of the limitations of the DF tools they use [5]. DRF guidelines such as [8], published by government agencies, and DRF case-studies such as [10] and [12] by academics, could assist DF investigators to address these challenges [13].

Forensic case studies of older drone models, particularly DJI, appear widely in the literature; however, there appears to be a gap relating to the newer DJI Mini 3 Pro and its remote controller (RC) the "DJI RC". This is not surprising given that both devices were only recently released to the market. This study aims to address this gap by performing a static forensic case study on these devices, with a particular focus on identifying the presence of key digital artefacts that could be useful to investigators. Additionally, the paper explored methods for visualising the drones flight history and assessed some of the potential anti-forensic (AF) methods that could be utilised on these devices. It is hoped that the findings of this case study will assist investigators and researchers in future DRF analysis of these specific devices.

This paper is structured as follows: Section 2 discusses related work in drone forensics. Section 3 explores the methodology applied in the case study. Section 4 presents the results and analysis. Section 5 discusses findings of the research, and lastly, Section 6 of the paper concludes with future research directions.

.

## II. BACKGROUND AND RELATED WORK

### A. Key Data Types

There are several types of digital artefacts potentially valuable to forensic investigators. According to Interpol [8], these include recorded media (digital imagery and video footage), flight logs, payload created content (e.g., drop zone locations), Personally Identifiable Information (PII), and usage logs (e.g., system and sensor logs). These key data types were the focus for drone forensic case studies completed by [3] - [5], [10], [14], and [15], further supporting the recommendations provided by [8].

While flight logs are key evidentiary items in DRF [8], extracting and processing them can be time-consuming, as they can be encrypted, granulated, and unstructured in their raw form [14]. Their structure can also vary significantly between drone makes and models, which can present further challenges for investigators [13], [14]. Case studies on the DJI Phantom 4, Yuneec Typhoon, and the Parrot Bebop by [14] found that the flight logs for the Parrot Bebop and Yuneec Typhoon were unencrypted and able to be parsed more easily than that the DJI Phantom 4, which in comparison had encrypted logs containing substantially more parameters.

Although payload data could indicate transportation of illegal goods by a drone, only [10] included it as a focus of their drone forensic case study. It appears that research focusing on this data type is lacking in current literature.

In terms of PII, [4], [10], and [13] mention attribution of drone ownership as a key goal for investigators. Some of the types of PII include the user account of the drone operator [8], their location data [7], and serial numbers of the drone, RC, and batteries [13]. Researchers [4] concede that PII data is more likely to be found on the RC rather than the drone itself.

### B. Data Sources

There are several possible sources for acquiring data from drones. The sources typically include non-volatile memory sources (internal memory and external memory cards), volatile memory sources (e.g., Random Access Memory), sensor data (e.g., propellers and camera), and wired and wireless network systems (e.g., Universal Serial Bus [USB], Wi-Fi, and Bluetooth) [9]. Studies by [3], [13-15], examined data acquired from non-volatile memory sources of the drone/RC, specifically the internal memory of the drone and smartphone-based RCs, and removable SD cards used in those devices. Conversely, the study by [7] focused on wireless data produced by the DJI's DroneID system (a proprietary wireless drone identity and location beaconing system) on several DJI drones, while [3] applied vulnerability testing exercises against two drone models in an attempt to examine wirelessly transmitted data over the drone's Wi-Fi systems. Drone forensic guidelines by [8] mention the following possible sources of data: removeable memory cards, internal memory, cloud-based storage, wireless systems (e.g. Wi-Fi and Bluetooth) and Internet services used by the drone.

While [5] and [11] suggest there is value in data gathered from wireless-based signals, [11] stressed that this type of evidence is considered less significant when compared to static forms of digital evidence.

### C. Acquisition and Analysis Approaches

As seen in [3], [7], and [15], drone forensic acquisition and analysis approaches can be grouped into two broad categories. One of these is the "static" approach, which involves extracting and analysing data at rest which is not subject to change [16]. An example of a static approach, one of which is demonstrated by [10], is extraction and analysis of metadata stored within images extracted from a Secure Digital (SD) card used in a drone. The static approach typically involves imaging a disk of the target system while it is powered off [17]. The other type of approach is a "live" approach, which involves extracting and/or analysing data that is dynamic and subject to change [16]. In the context of DRF, this would include wireless data transmitted by the drone's systems and interacting with the devices while they are powered on [3]. Live approaches were taken by Schiller et al. [7] and Salamh et al. [3]. Of the two approaches, the static approach was most widely used among studies by [4], [5], [10], and [12] - [15]. This may be due to the additional complexities often involved in live forensic approaches.

There are four data extraction methods that can be applied to DRF: manual extraction, logical extraction, physical extraction, and wireless extraction. Manual extraction involves direct interaction with the remote controller and/or taking photos of what is shown on the screen, while logical extraction obtains data from a readily available part of the file system (usually presented as directories and files and does not normally include access to unallocated space nor previously deleted files) [18]. Physical extraction utilises a bit-by-bit copy of the storage medium, and, unlike logical extraction, includes unallocated space [9], [18].

Physical extraction methods may include reading the data from a chip located on its host device [15], and are usually performed on a wired-interface such as a Universal Asynchronous Receiver Transmitter (UART) [7], or Joint Task Action Group [8], also located on the host device. Case studies by [3], [4], [12], and [13] successfully retrieved digital artefacts using some or a combination of these methods.

Other physical extraction methods include 'chip-off', 'micro-read' and capturing wireless Radiofrequency (Rf) signals [19]. The 'chip-off' method involves removing the target chip from its host device and placing it in a reader or another similar host device to extract the data directly from the chip [18]. Unless the chip has full disk encryption, this method could potentially bypass any authentication and encryption mechanisms applied by the chips' usual host device [18]. The 'micro-read' method involves removing the top layers of silicon from the storage chip to expose the logic gates so they can be examined with electron microscope which can read the data [9]. Wireless data extraction methods typically involve the use of a wireless Rf receiver, such as the Software Defined Radio (SDR) receiver used by [7] to capture wireless Rf signals from several DJI drone models.

Both the 'chip-off' and 'micro-read' methods typically offer investigators the most amount of data from the drones, however they are more complicated to undertake, require specialist hardware and tools, and are often destructive to their host target devices [4]. Researchers in [13] had success in partially recovering a previously deleted image using a chip-off method on a DJI Phantom 4 and [15] also recovered data from a DJI Mavic Air 2 using this method. This data was not discovered by the less invasive methods. However, utilising extraction methods that are likely to alter or damage the original source of data or host device is not best practice in DF and is only recommended when data deletion is suspected and/or data retrieval by other methods has been unsuccessful [9], [13].

Drone forensic guidelines by [8] recommend that investigators apply a variety of acquisition and analysis methods to drone investigations. This recommendation is supported by [10].

Another approach to gaining access to normally inaccessible areas of a storage device is to obtain root privileges [15]. This is usually achieved by exploiting a vulnerability in the software running on the device [15]. Studies by [3] and Schiller et al. [7] involved the use of vulnerability testing techniques in order to gain elevated privileges and access data from the drone/RC that were normally off-limits to the user. Although [7] identified methods for gaining root access to an RC used in their study, they did not publicly disclose the details of the vulnerabilities.

Although methods to gain elevated privileges could yield promising results, it could unintentionally alter the data on the target device, which would be a violation of one of the Association of Chief Police Officers (ACPO) principles of DF [20, p 6,],"Principle 1: No action taken by law enforcement agencies, persons employed within those agencies or their agents should change data which may subsequently be relied upon in court".

To generally avoid altering data on a target device, it is recommended to use a write blocking device where possible during the data extraction phase [8], [17]. Existing DRF case studies by [5], [10], [15] mentioned using either hardware- or software-based write blockers when imaging data from SD cards.

Some DF studies have focused on drone wireless systems. For example, [7] examined DJI's DroneID system using a SDR receiver. Their study concluded that the data transmitted by the version of DroneID operating on several DJI drone models examined in their study was not encrypted, despite the manufacturers claims otherwise [7]. Schiller et al. [7] were also able to manipulate the DroneID system to send a spoofed serial number and GPS location of the RC.

DJI have also developed Ocusync, a proprietary wireless system for control and video downlink of compatible DJI model drones [7]. The DJI Mini 3 Pro and DJI RC both reportedly use Ocusync version 3 [21]. Like DroneID, the specifics of this technology have not been publicly documented by DJI [7]. Schiller et al. [7] study also examined Ocusync 2.0, which they believed was the host transmission protocol for DroneID.

*D. Key Challenges*

Although attribution is a key goal of DRF, it can be difficult to achieve for various reasons. [14] cited technical reasons as a significant challenge. As an example of this, [10] unintentionally found that the DJI Mini 2 drone examined in their study was unable to get a GPS signal lock and fetch location coordinates when it was flown inside a sports hall. While this finding was the result of a technical limitation, [10] also highlighted it as a possible AF technique that could be exploited.

To help address the difficulties associated with tracking the identity and location of drones operating in United States (US) airspace, the FAA will be mandating a wireless identity and location broadcasting system, which they are calling RemoteID, by the end of September 2023 [22]. Their mandate stipulates that all drones operating in US airspace will need to be wirelessly broadcasting their identity and location to authorities, and that all drone pilots will need to be registered with the FAA [22]. To help DJI drone owners meet the FAA's RemoteID requirements, DJI have implemented their internally developed RemoteID system (DroneID) on several of its drone models [7]. Despite this, a study of DJI's DroneID system by [7] found that it was susceptible to spoofing, which could lead to misattribution of a drone operator and/or complete avoidance of attribution.

There are a variety of tools available for a DRF investigation, and often several tools are required [10]. For example, a case study of the DJI Mini 2 by [10] encountered issues when they attempted to decrypt .DAT type flight logs using DatCon, but were able to successfully decrypt them in an online flight log parsing and visualisation site called "Airdata UAV" (https://app.airdata.com/).

Another key challenge in DRF relates to limitations of forensic tools. The tools available to a forensic investigator could turn out to be unreliable for a variety of reasons. For example, DatCon, an open-source flight log decryption tool was successfully used by [3], [13] to decrypt flight logs acquired from a DJI Phantom and a DJI Matrice, however this tool was reportedly not able to decrypt flight logs from a DJI Mini 2 when attempted by [10]. The inability of the tool to be able to decrypt the logs from the Mini 2 is likely due to changes made in the encryption and/or specification of the flight log standard, which the manufacturer can change at any time in a firmware update. In another example, [3] assessed the capabilities of two popular forensic analysis tools: Autopsy and Cellebrite, and found a slight discrepancy between the timestamps presented for a flight log acquired from a DJI Phantom. The researchers in [3] also discovered that Autopsy and Cellebrite generalised the flight log waypoints, while DatCon produced the most accurate results. However, the same researchers also found slight differences in the values outputted by log files parsed by DatCon when Java compiler was used. Slight discrepancies can have significant forensic implications [3]. To this end, researchers [4], [5], and [10] recommend that investigators have a thorough understanding of the capabilities and limits of the available tools.

A lack of DF methodologies and guidelines pose another key challenge. According to current literature, there is no consensus among researchers on methodologies, guidelines, and standards for DRF investigation. For example, researchers [10] referenced application of the National Standards and Technology (NIST) DF guidelines for mobile devices [19] for their case study of the DJI Mini 2, while [9] and [13] cited Interpol DRF guidelines as suitable.

Researchers [9] and [11] reviewed several existing DRF models and identified gaps in the models. [9] claimed that existing DRF methodologies focus on traditional computer systems or mobile devices, which were not specific enough to address the cyber-physical nature of drones, while [11] posited that many of the existing models were not versatile enough to provide coverage for all drone models and technologies. Both [9] and [11] agreed that the DRF field was in need of standardised international guidelines. In an effort to further develop an international standard, researchers [11] proposed their own model which they believed made up for the shortcomings of the models they reviewed. Their proposed model considered both legal and technical aspects of the DRF domain and included a preparation phase, something they claimed was lacking in existing models. Despite claims by [9] and [11], Interpol, an international organisation, had already made efforts to develop an international standard by publishing their own DRF guidelines [8].

The ability to identify and track drones and their pilots is another challenge of DRF. RemoteID, which the FAA is mandating in September 2023, is envisioned to assist forensic investigators by providing authorities with drone tracking and identification of pilots in US airspace [22].

*E. Gaps in the Literature*

The majority of forensic case studies in the literature focus on DJI models. The reason for this is likely due to the fact that DJI are the most popular drone manufacturer, claiming a 76% global market share [23]. To best of authors knowledge, there are no published DRF case studies on the Mini 3 Pro or the DJI RC.

In addition, [10] claimed there was a lack of literature pertaining to DRF of drones weighing less than 250 grams. Their study of a DJI Mini 2 attempted to partially address that gap.

The majority of DRF studies reviewed that involve flight log analysis, used online based tools, which require uploading the data to a third-party provider. This may raise privacy concerns for sensitive cases. None of the literature reviewed appears to have utilised the Flight Reader software, a Windows-based application that was designed to locally decrypt and present DJI flight logs in text and visual forms.

A review of the literature indicates a lack of research into AF methods for drones. This conclusion is echoed by [4], who calls for more research into AF techniques of drones. "CIAJeepDoors" is an open-source Python program that its author claims is capable of disabling DroneID on some DJI drones [24]. This application was not mentioned in any of the reviewed literature.

III. METHODOLOGY

The selected research methodology in this paper aimed to address the research questions presented in Section 1 by a) comprehensively assessing the digital evidence characteristics of the DJI Mini 3 Pro and DJI RC, b) extracting and analysing key digital artefacts from the selected devices, c) exploring methods for visualising the drone's flight history, and d) highlighting potential anti-forensic tactics for these devices. An individual case study design was selected to address the research questions. The parameters of this case study are described in further detail below.

*A. Experiment Design*

To minimise the amount of unrelated data and cross-contamination of data in the case study, the experiment environment was controlled as much as possible given the resources and time available during the study.

To keep the study within key data of interest, data acquisition and analysis methods were aligned with guidelines by [8] *"Framework for Responding to a Drone Incident: For First Responders and Digital Forensics Practitioners"*.

In addition, four scenarios were devised to simulate possible real-life situations initially faced by a digital forensics' investigator performing a drone forensic investigation. These scenarios were:

*1) Scenario A:* In this scenario, only the drone is recovered by authorities at the scene. The device is fully functional and contains an external Micro SD card. The RC did not contain an SD card prior to, or during, the flight. The flight involved a safe take-off and landing and was unremarkable.

*2) Scenario B:* In this scenario, only the RC controller is recovered by authorities at the scene. The device is fully functional and contains an external Micro SD card which was used during the flight. Additionally, the SD card in the RC was selected as the primary storage device and the drone did not contain an SD card. The flight involved a safe take-off and landing and was unremarkable.

*3) Scenario C:* In this scenario, both the drone and RC are recovered, they are fully functional, and both contained an SD card during flight. The flight involved a safe take-off and landing and was unremarkable.

*4) Scenario D:* In this scenario, both the drone and RC are recovered, however neither device contained an external SD card during the flight. Additionally, the drone was flipped upside down by hand toward the end of the flight to simulate a collision. The simulated collision resulted in both devices being partially damaged prior to the investigation, and as a result, the Universal Serial Bus (USB) ports were inoperable during the investigation. Despite this, both Wi-Fi and Bluetooth systems were functional. Furthermore, to simulate a potential anti-forensic tactic, the user account was logged out of the DJI Fly app on the RC immediately prior to the flight.

These scenarios were conducted in the order presented. In all scenarios, the identity of the drone's operator and usage are treated as unknown, and the goal of the forensic investigator is to try to determine this information from digital artefacts acquired from the devices.

This DRF case study also identified possible digital anti-forensic methods available for these devices. However, due to time constraints, these methods were not thoroughly tested during the study.

The primary focus for this DRF case study was on the digital data stored on the internal storage memory of the selected devices and removable storage devices used during the experiment. DJI's cloud storage service and proprietary wireless systems, Ocusync and DroneID, were not in scope of the study.

*B. Process workflow*

The high-level workflow for the practical elements of the case-study is shown in Figure 1. Each scenario starts with a preparation phase, followed by data generation, then extraction and analysis, and lastly, recording of the results.

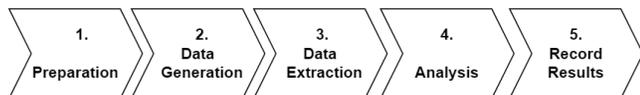

Fig. 1. The generic steps followed for practical elements of the case study.

Further detail on the preparation, data generation, extraction, and analysis phases are provided in the following sections. The 'record results' phase entails making comprehensive notes of the results and documenting them in this article.

*C. Preparation and Equipment*

*1) Forensic Workstation Setup:* A laptop was set up purposely for the experiment. It comprised of an i9 Intel Core processor, 64 GB of ram, and a 1 TB hard drive. The host OS of the machine was Windows 11. Virtual machine (VM) environments were setup specifically for the experiment. The VM's were Windows 10 OS (see Microsoft.com) and SANS SIFT Linux Workstation (see sans.org).

*2) Tools:* Several tools were utilised in the case study. Table 1 outlines these tools. The primary digital forensic tools used in the study (FTK Imager and Autopsy) are popular in the digital forensics field. Two DJI developed applications, DJI Fly and DJI Assistant 2, were used in the study. DJI Fly was utilised for a range of activities, including controlling the drone, displaying the drone's live video feed on the RC, capturing, and viewing recorded media, and transferring recorded media between the RC and a compatible device. The DJI Assistant 2 tool is a Windows/Linux-based tool that provides the ability to extract logs from non-user accessible storage areas of the Mini 3 Pro and DJI RC. These tools assisted in examining and extracting digital artefacts devices from the devices.

Several tools were used for parsing and visualising flight logs over satellite imagery. DatCon and Flight Reader (https://www.flightreader.com/) were used for initial decoding and parsing of the logs, while the 'Airdata UAV' website, Google Earth Pro, and Flight Reader were utilised for flight path visualisation.

Many of the tools selected for the study were either freeware, open-source, or already owned and licenced by the researcher prior to the study. Open source and freeware tools were preferred over proprietary tools for this case study due to cost constraints. The tools used in the study were installed on their respective VM's as outlined in Table 1.

*3) External Storage Cards:* Two SanDisk Ultra Class 1 Micro-SD cards (a 16 GB and 32 GB) were used in scenarios A, B and C. To lessen the possibility of cross-contamination of data between the scenarios, both Micro SD cards underwent data sanitisation prior to and in between the data generation phases for scenarios A, B and C. The SD Association's card formatting tool "SD Card Formatter" was used to prepare both cards prior to generating data for Scenario A; however, it was discovered during subsequent analysis of this scenario that previously deleted files were still present on the card, leading to cross-contamination of data between the scenarios. For successive scenarios, the native Linux Data Duplicator utility was used to zero out the SD cards. Additionally, after each sanitisation activity, the cards were carefully examined using Autopsy to verify that they did not contain any remnants of data. When it was confirmed that a card did not contain data, the SD card for the drone was formatted using the native format tool in the Windows OS and the card used in the DJI remote controller was formatted using the native Android Files utility on the RC.

*4) Mini 3 Pro Drone:* The DJI Mini 3 Pro (Model: MT3M3VD) is a sub-250g multi-rotor consumer class drone developed by DJI. Its key features are a 3-way collision avoidance system, GPS connectivity, a stabilized 3-axis gimbal, and a 1/1.3-inch sensor camera that is capable of recording 4K resolution video and capturing 48-megapixel images [21]. A topside view of the DJI Mini 3 Pro drone that was examined in the study is shown in Figure 2.

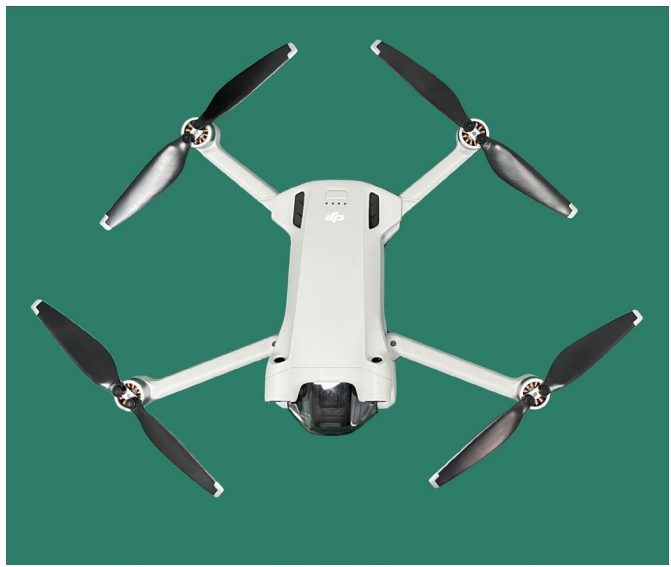

Fig. 2. A top-down image of the DJI Mini 3 Pro drone.

The drone and RC utilise DJI's Ocusync protocol (version 3.0) for control plane and video feed communication [21]. Currently there is no option for a user to factory reset the drone or format or wipe its internal memory. The only known non-invasive option available to users is to delete existing files and directories from the drone's internal drive while it is mounted by a suitable device. This is not an effective method for sanitising data on the drone's internal storage. As the drone was extensively used prior to the experiment, it contained data from previous usage. This limitation was factored into the experiment during the extraction and analysis phases, and as such, data identified as having been generated prior to, or outside of the experiment, was unless specifically mentioned in the findings, deemed out of scope. Preparation for the drone included ensuring that it was fully charged prior to and during the data generation, extraction, and analysis phases, calibrating the drone with the RC-based version of the DJI Fly application, performing visual checks on the drone body for signs of damage, and placing the rotor arms into flight position prior to flight.

TABLE 1
HARDWARE AND TOOLS USED

| Hardware/Tool | Version | Description | Usage | Availability | Vendor Link |
|---|---|---|---|---|---|
| *Flight system* | | | | | |
| DJI Mini 3 Pro | Model: MT3M3VD. Firmware 01.00.0400 | Drone. | Generate data and examination target | N/A | www.dji.com |
| DJI RC | Model: RM330. Firmware 01.02.0100 | Drone remote controller. | Drone navigation/run DJI Fly application/capture live video feed and examination target | N/A | www.dji.com |
| DJI Fly | 1.8.0 (RC) 1.9.1 (iPhone 13) | Support tool | Drone control/Data transfer | Freeware | https://www.dji.com/downloads/djiapp/dji-fly |
| SanDisk Ultra Micro SD card | 32 GB SDXC Class 1 | Data Storage. | External data storage for drone | N/A | sandisk.com |
| SanDisk Ultra Micro SD card | 16 GB SDXC Class 1 | Data Storage. | External data storage for RC | N/A | sandisk.com |
| *Forensic extraction and analysis tools* | | | | | |
| Forensic workstation | Metabox Prime-SR | Windows-OS-based laptop | Primary forensic workstation | Proprietary | metabox.com.au |
| SIFT Ubuntu | 20.04 | Linux-OS | VM environment | Open source | https://www.sans.org/tools/sift-workstation/ |
| Windows 10 | 10 | Windows-OS | VM environment | Proprietary | https://www.microsoft.com/en-au/software-download/windows10 |
| VMware Workstation Pro | 16.2 | VM application | For running VM environment on host OS | Proprietary | vmware.com |
| Apple iPhone | 13 | Smartphone based DJI Fly app installed | To acquire media and flight logs via Wi-Fi/Bluetooth | N/A | apple.com |
| DJI Assistant 2 (consumer series) | 2.1.15 | Support tool. | Firmware update/Data download | Freeware | https://www.dji.com/downloads/softwares/dji-assistant-2-consumer-drones-series |
| Linux Data Duplication Utility | N/A | For wiping data on MicroSD cards | Preparation for data generation | Open source | linux.org |
| Micro SD card adapter | N/A | With hardware write blocker switch | hardware based write blocker | N/A | www.sandisk.com |
| SD card reader | N/A | Unitek USB card reader | Data extraction | N/A | www.unitek-products.com |
| Autopsy | 4.19.3 - Windows 64-bit | Forensic imaging and analysis tool | Imaging/hashing/analysis | Open source | https://www.autopsy.com/download/ |
| AccessData FTK Imager | 4.3.0.18 – Windows 64-bit | Forensic imaging and analysis tool | Imaging/hashing/analysis | Freeware | https://www.exterro.com/ |
| Exif Tool | 5.16.0.0 - Windows 64-bit | Image metadata viewer | Extract metadata from images | Freeware | http://u88.n24.queensu.ca/exiftool/forum |
| VLC Media Player | 3.0.18 - Windows 64-bit | Video viewer | View recorded videos and flight telemetry | Freeware | https://www.videolan.org/vlc/ |
| DatCon | 4.2.6 - Windows 64-bit | Offline DJI Flight log decrypting tool | Alternative tool for decrypting DJI logs | Open source | https://datfile.net/DatCon/intro.html |
| Notepad++ | 7.9.5 - Windows 64-bit | Text editor | For viewing .DAT and .TXT files | Freeware | https://notepad-plus-plus.org/ |
| Infranview | 4.62 - Windows 64-bit | Image viewer | For viewing Joint Photographic Experts Group (.JPG/.JPEG) images | Freeware | infranview.com |
| Infranview Plugins | 4.62 - Windows 64-bit | To view additional image filetypes | For viewing Digital Negative (.DNG) images | Freeware | infranview.com |
| HxD Hex Editor | 2.4.0.0 – Windows 64-bit | Hex editor/viewer | For viewing files at bit level | Freeware | www.mh-nexus.de |
| Flight Reader | 1.4.6 - Windows 64-bit | Offline DJI Flight log analyser | Decrypt flight logs/visual analysis of flight data | Proprietary | https://www.flightreader.com/ |
| Google Earth Pro (Desktop version) | 7.3 - Windows 64-Bit | Satellite map viewer | To visualise the decrypted log files. | Freeware | google.com/earth |

*5) DJI RC:* The DJI RC (Model RM330) is an all-in-one remote controller that is compatible with the DJI Mini 3 Pro drone [21]. It features an integrated touchscreen for viewing live video feed from the drone and recorded videos and images. It also features controls to operate the drone, USB ports for connecting the RC to external devices, an 8 GB internal memory storage device, and a Micro-SD card slot for extending storage capacity [21]. Preliminary examination of the device indicated that the device's host OS was Android version 10. A topside view of the DJI RC that was examined in the study is shown in Figure 3.

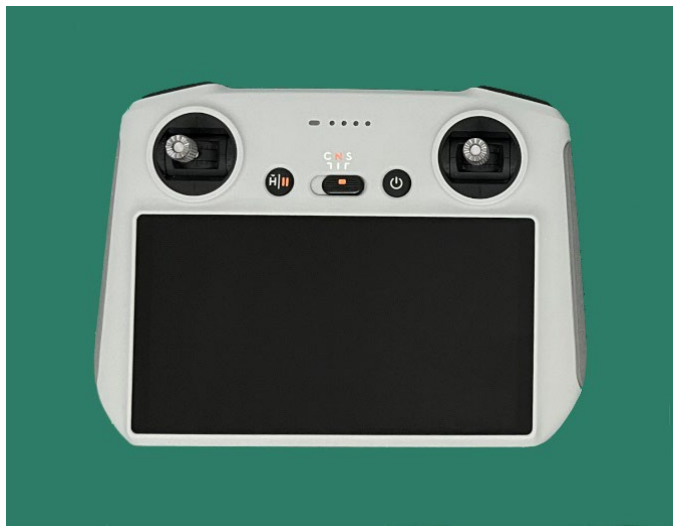

Fig. 3. A top-down image of the DJI RC.

The DJI RC was shipped with the DJI Fly application pre-installed. The controller was prepared and configured for the experiment as follows:

- It was paired with the Mini 3 Pro drone used in the study.
- A DJI user account was logged in to the DJI Fly app during Scenarios A, B and C.
- The DJI user account in the DJI Fly app was bound to the drone during the study.
- DJI Fly app privacy settings:
    - "Local Data Mode" was disabled.
    - "Mobile Device GPS Info" was disabled.
    - "DJI Device Hardware Info" was disabled for some of the scenarios.
    - "Approximate Location info" was enabled.
- The Wi-Fi option on the Android OS was set to off.
- Video subtitles were enabled.
- The RC was not jailbroken/rooted.

*6) Apple iPhone:* A smartphone version of DJI Fly application is also available for compatible Apple iOS devices from the App Store [25]. When the application is paired with a compatible smartphone-based DJI remote controller (such as the model RC-N1) it can act as a flight controller and provide similar features to the DJI RC. However, the smartphone application can also connect directly to the DJI RC or the Mini 3 Pro drone via the Bluetooth and Wi-Fi protocols to transfer multimedia to the smartphone. To provide a wireless data extraction method for Scenario D, the DJI Fly application was installed on a used Apple iPhone 13 and paired with the Mini 3 Pro Drone and DJI RC prior to the experiment. Additionally, the researchers' DJI user account was logged into the iOS DJI Fly app prior to, and during, the experiment.

## D. Data generation

To generate sufficient data for the practical elements of the case study, at least one flight of the drone was undertaken for each of the scenarios outlined in Section III-A. Additionally, each flight was controlled using the DJI RC and at least one video and image were taken during each flight.

## E. Acquisition

Data acquisition guidelines outlined in [8] were adapted for this case study. The primary preparation and acquisition processes for the scenarios of this case study are outlined in Figure 5.

This study was limited to manual, physical, and logical extraction methods and where possible, in this order to conduct the examination in a forensically sound manner as recommended in [8].

Data integrity has also been considered. To meet the minimum data integrity requirements, image files created by FTK Imager were hashed, and a hash report generated for the image.

Specific extraction methods are detailed as follows:

*1) Mini 3 Pro drone:* The Mini 3 Pro drone features a 1.2 GB capacity internal storage device, a Micro SD card slot for external storage, and a USB-C port for connecting to external devices. Figure 6 show the port placement in respect to the drone.

As per the process workflow depicted in Figure 5, if the scenario involved use of a Micro-SD card in the drone, then this was imaged prior to attempting data extraction from the drone's internal memory. Methods used for extracting data from the drone's internal memory included:

- Connecting the drone to the forensic environment via a wired USB connection and exploring the user accessible volume presented to the forensic VM (i.e., a logical acquisition).

- Using the DJI Assistant 2 tool (via a wired USB connection) to access areas of the drone's storage not presented to the forensic VM.
- Using the DJI Fly app, on both the DJI RC and the Apple iPhone, to wirelessly connect to the drone and extract multimedia files from it.

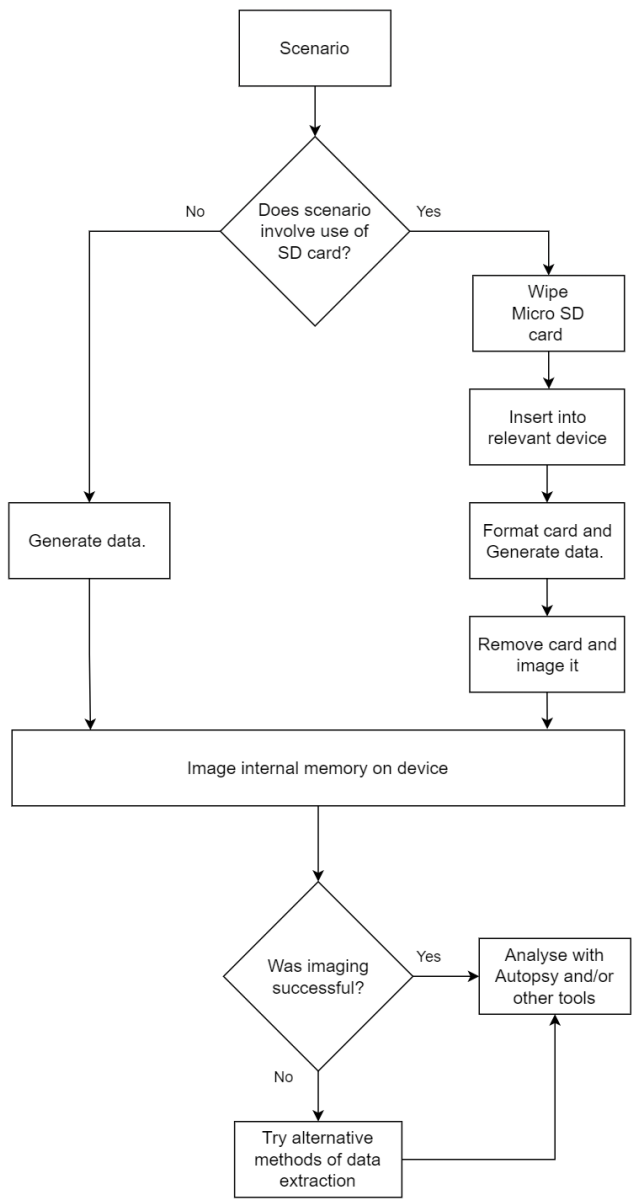

Fig. 5. The preparation and acquisition processes adopted in this case study.

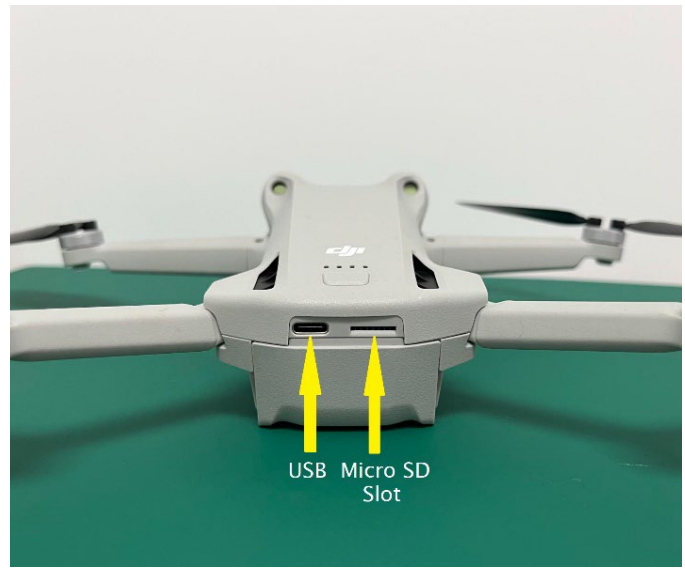

Fig. 6. A view of the USB port and Micro-SD card slot on the Mini 3 Pro.

*2) DJI RC:* The DJI RC features an 8 GB capacity internal storage, a Micro SD card slot for external storage, and two USB-C ports for connecting to external devices. Figure 7 shows the port placement on the RC.

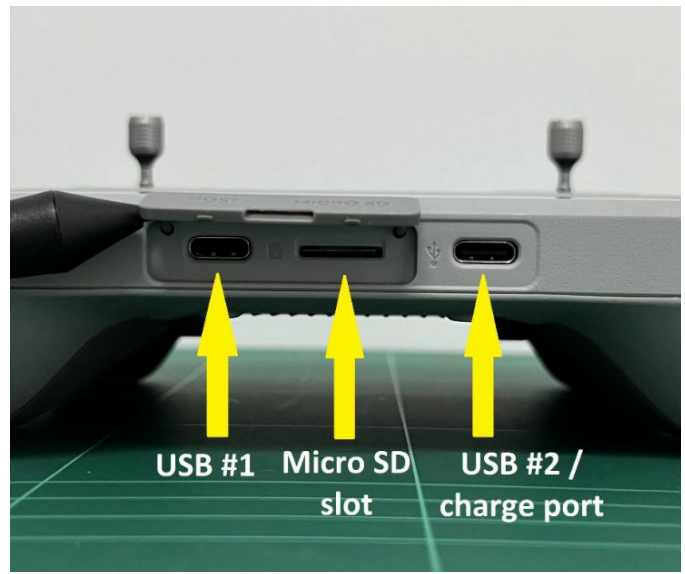

Fig. 7. USB ports and Micro-SD card slot location on RC.

Similar data extraction methods carried out on the drone were adopted for the RC, however, as the RC is an integrated device, manual extraction methods were an additional option that were explored. The Micro-SD cards used in Scenarios A, B and C were imaged before other possible extraction methods (i.e., manual, and logical) were explored on the RC's internal memory. The DJI Assistant 2 tool was used to attempt to extract flight and system logs from the RC while Windows Explorer and Linux Files utility on the RC's host-OS and the Linux forensic VM were used to explore the RC's internal memory.

*3) External Storage:* A 16 GB SanDisk Ultra High-Capacity Micro SD card was present in the RC during the data

generation phase of Scenarios B and C while a 32 GB SanDisk Ultra High-Capacity Micro SD card was present in the drone during the data generation phase of Scenarios A and C.

A SanDisk Micro-SD to SD card adapter was used as a hardware write blocking device while imaging the Micro-SD cards. Prior to imaging a Micro-SD card, the adapters' write block switch was placed into the enabled position (as shown by the red arrow in Figure 8). Placing the switch into this position blocks write operations on the Micro-SD card **[26]**. The write block switch was placed into the disabled position for data sanitisation activities.

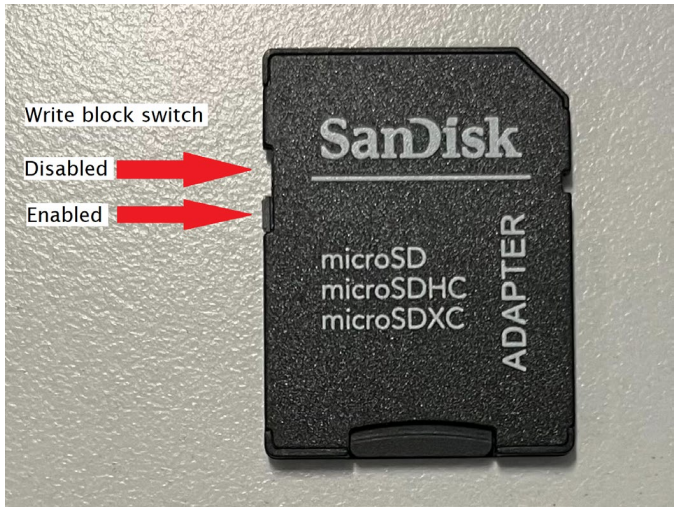

Fig 8. Micro-SD card adapter with write block switch enabled.

*F. Analysis*

Data was analysed using the tools outlined in Table 1. The general processes that were followed for the analysis are depicted in Figure 9. The primary analysis tools were Autopsy, HxD, the EXIF Tool, Notepad++, DatCon, and Flight Reader.

Specific analysis steps were undertaken for suspected flight logs. This included submitting the files to both DatCon and Flight Reader for initial decoding and parsing. DatCon did not have an integrated visualisation feature; therefore, logs parsed by DatCon were uploaded to the Airdata UAV and Google Earth for visualisation analysis. These steps are depicted in Figure 10.

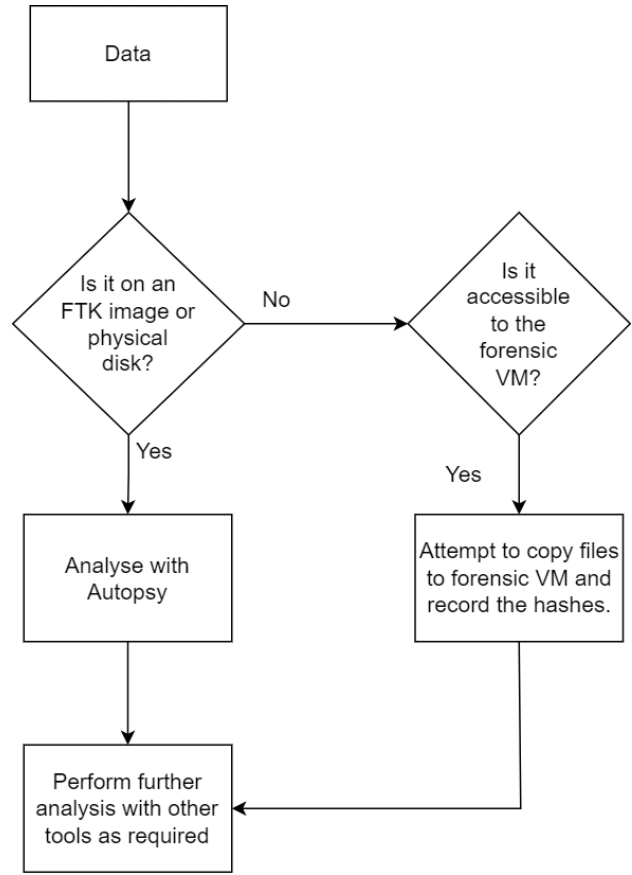

Fig. 9. General data analysis processes

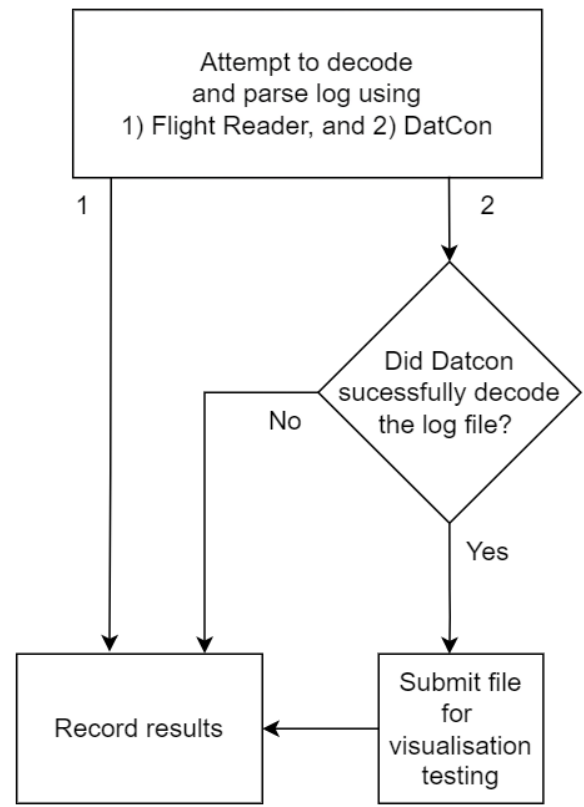

Fig. 10. Flight log parsing and visualisation processes

## IV. Findings

Digital artefacts were discovered and retrieved from the Mini 3 Pro and DJI RC. These were grouped into three primary categories: media, flight logs, and other. The findings of the scenarios are detailed in Subsections A to D. Additionally, anti-forensic methods were identified; the results of which are outlined in Subsection E. A summary of the overall findings is provided in Subsection F.

### A. Scenario A

For this scenario, analysis was first performed on the FTK-acquired image of the 32 GB Micro-SD card that was used in the drone. This was followed by analysis of the drone's internal memory. Autopsy was used in both cases.

*1) SD Card:*

*a) File System/Structure*

- Filesystem type was FAT32.
- Three parent folders were present in the root volume:
  - DCIM
  - LOST.DIR
  - MISC

*b) Media*

Images and videos were discovered in several locations on the SD card image, including unallocated areas.

- High-resolution images were successfully retrieved from folder "DCIM\100MEDIA". These included raw (.DNG) and compressed JPEG (.JPG) versions of each image captured by the drone.
- A high-resolution video taken for Scenario A was retrieved (in full) from "DCIM\100MEDIA". The filename was DJI_0161.MP4. The video was able to be replayed in full using VLC Player.
- A pair of thumbnail images of the first frame of the recorded video was discovered in the "MISC\THM\100\" folder. The filetypes of these images were .SCR and .THM. However, a hex analysis of these files revealed that they were both JPEG files (see Figures 11 and 12). The image files ending in .THM were 160 x 90 pixels wide while .SCR files were 960 x 540 pixels wide. The filename of these images corresponded with the filename of the related video ("DJI_0161.MP4").
- Previously deleted images were discovered by Autopsy in an unallocated portion of the SD card. These images were a mixture of JPG and DNG type files. The recovered JPEG images were thumbnails of older images captured by the drone prior to the case study. The pixel size of these recovered images matched the respective pixel sizes of the .THM and .SCR files. The DNG images recovered by Autopsy were 48-mega pixel high-resolution images captured by the drone prior to the case study. The creation timestamps of the previously deleted images were not the original creation date of the images, rather the timestamp was the point in time when the images were extracted by Autopsy.
- A subtitle file (DJI_0161.SRT) was present in "DCIM\100\MEDIA" folder. The file was viewed in notepad++ for further analysis. It appears this file contains metadata of the flight of the related video (DJI_0161.MP4). As shown in Figure 13, it was discovered that when subtitles are enabled and used in conjunction with the related video in VLC player, flight telemetry data is displayed on screen during playback.

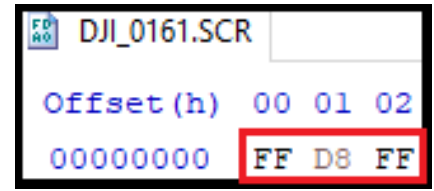

Fig. 11. The magic number of DJI_0161.SCR.

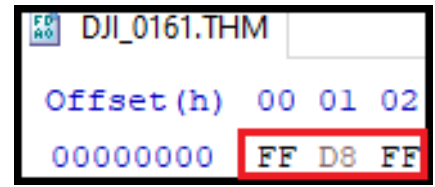

Fig. 12. The magic number of DJI_0161.THM.

*c) Flight Logs*

No fight logs were discovered on the Micro SD card image.

*d) Other*

- Autopsy discovered a total of 236 deleted files on the unallocated volume of the SD card. These included images (in DNG and .JPG format), movies (in .MOV format), an Adobe flash file (.SWF) and flight logs (in .TXT format). Of these files, the movie and flash files were 0 bytes and could not be recovered.
- Of the total recovered from the unallocated portion of the SD card, 131 of them were .txt type files. These files had filenames in the format: "fxxxxxx.txt" ("x" representing a number value). Closer inspection of these files with Notepad++ revealed they contained the following flight history metadata:
  - Date/time of flight
- GPS coordinates (latitude, longitude, and altitude)
- Camera parameters (such as shutter speed, iso, zoom ratio, and sensor temperature).

The type of data and format of data in these files was similar to that observed in DJI-generated subtitle telemetry files. It is believed these text files were previously deleted subtitle files. A snippet of this data is highlighted in Figure 14. Note that the latitude and longitude coordinate values have been partially redacted.

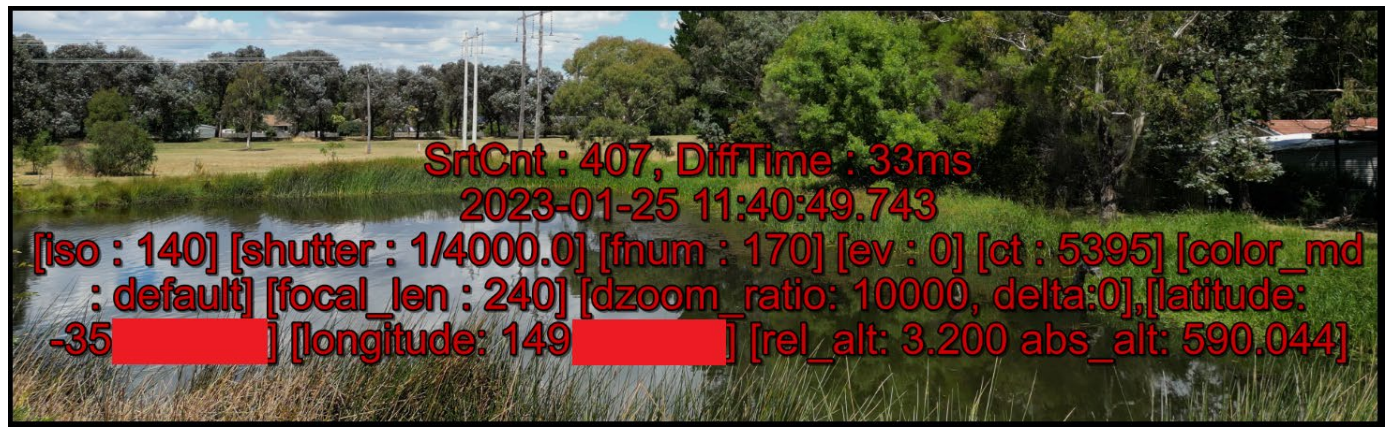

Fig. 13. Screenshot of video overlay telemetry file showing camera data. Note that the GPS coordinates have been partially redacted.

Fig. 14. Snippet of the contents of a text file recovered by Autopsy from an unallocated space of the Micro-SD card. Note that the highlighted GPS coordinates have been partially redacted.

- The images recovered from the SD card were analysed for the presence of Exchangeable image file format (EXIF) metadata. It was discovered that both the compressed JPEG (.JPG) and raw images (.DNG) located in "DCIM\100MEDIA", and the raw (.DNG) images discovered in and recovered from the unallocated volume contained comprehensive EXIF metadata. The information stored included timestamps of when the images were created, the make and model of the drone, camera sensor information (including its serial number), and GPS coordinates of where the image was taken. Spot checks of the EXIF metadata confirmed it was accurate.

*2) Internal Memory Analysis:*

   *a)    File System/Structure*
- Filesystem type was ExFAT.
- 1.5 GB user-accessible volume
- Mounted in Windows as a Mass Storage Device.

*b)    Media*

Nil found.

*c)    Flight Logs*

Nil found.

*d)    Other*

- A .DAT file was retrieved using DJI Assistant 2. The file was saved to the forensic VM as "DJI_Mini_3_Pro_2023-01-25_21-38-00.DAT". It appears that the timestamp in the filename coincides when the file was extracted, not when it was originally created in the drone. Analysis of this file with Notepad++ indicated that it was a binary file. DAT files contain serials. Decoding was attempted with DatCon; however, the output files were either 0 or 1 KB in size and did not contain any useful information.
- DJI Assistant 2 provides an option to extract software system logs (called "Assistant logs"). Each extraction with this option generates two text files. The filename format for one of these files is "YYYY_MM_DD@HH_MM_SS.log" while the other is titled "ui_ass2.log. The log file with time and date in the filename appears to be in a mix of Korean and Chinese simplified encoded text. Online translation sites Google Translate (translate.google.com) and DeepL (deepl.com) were not able to decode it into anything meaningful. The other log "ui_ass2.log" contains multiple lines with timestamps in English, accompanied by encoded text. Text appears to be a Base64 or Base85 type encoding scheme. Attempted to further decode with various Cyberchef modules (cyberchef.org), but the attempts were unsuccessful. It is believed that both of these files were using a proprietary encoding scheme.

B.  Scenario B

In this scenario, only the RC was retrieved. The RC contained an SD card which was imaged and analysed using Autopsy. Manual and logical extraction methods using Windows Explorer and DJI Assistant 2 were explored for analysis of the RC's internal memory.

*1) SD Card Analysis:*

*a)    File System/Structure*

- Filesystem type was ExFAT.
- Several directories discovered in the root volume.

*b)    Media*

- No images were found on the SD card.
- The original video files of the recordings made during the flight were not found on the Micro SD card image.

- A cached video of a video recording taken during the flight for Scenario B was successfully recovered using Autopsy. The file was called "2023.01.28.19.09_49_Cache.mp4" and was located at: "/Android/data/dji.go.v5/files/MediaCaches/". The video was extracted to the forensics workstation for further analysis. The specifications of this video were:
  - o MPEG-4 file type
  - o 864x480 pixels
  - o Length: 00:01:38 h/m/s
  - o Creation date: 28.1.2023
  - o Creation time: 1911 hours
  - o Size: 49.5 MB
- While not technically part of this scenario, the related full-sized video was discovered on the drone's internal memory. The specifications of this video were:
  - o MPEG-4 file type
  - o 1920 x 1080 pixels
  - o DJI_0005.MP4
  - o Creation Date: 28.1.2023
  - o Creation Time: 1911 hours
  - o Size 437 MB

The cached version discovered on the RC was around 11% the size of the larger video discovered on the drone.

*c)    Flight Logs*

Nil found.

*d)    Other*

Several interesting patterns emerged when Autopsy's timeline feature was used on the SD card image.

- To begin with, it was discovered that several files and top-level directories were generated on the SD card around the same time (See Figure 15). The file creation timestamps reported by Autopsy correlated with the date and time the SD card was last formatted by the RC.
- The timestamp for a "file accessed" event entry for "music_sound_wave" coincided with the time the RC was booted up (see Figure 16). The associated sound file is played when the RC is powered on.

*2) Internal Memory Analysis:*

*a)    File System/Structure*

- Filesystem was presented to Linux and Windows OS as "Generic Hierarchical" type.
- 3.84 GB user accessible volume.
- The device mounted as a Media Transfer Protocol device, and as a result, imaging the drive was not possible with the available tools.

*b)    Media*

- Nil found via the volume presented to the forensic OS.
- Several thumbnail-sized images were discovered on the RC's internal memory using manual extraction methods. The images were located at "DJI RC\Android\data\dji.go.v5\cache\ImageCaches". (See Figure 17).
- It appeared that a pair of different sized thumbnails was generated for each recorded photo/video:
  - Filename "Photo_x_DJI_X_jpg_xxxxxxx_0_YYYYMDDxxxxx_photo_thunmbnail.jpg". (The word "thumbnails" were misspelt in these file types).
    - Size = 160 x 120 pixels
  - Filename "Photo_xxxxxxxxxx_DJI_XXX_jpg_xxxxxxx_0_YYYYMDDxxxxx_photo_preview.jpg".
    - Size = 960 x 720 pixels
  - Filename "Video_xxxxxxx_DJI_XXX_mp4_xxxxxxxxxx_xxxxxx_YYYYMDDxxxxxx_photo_thunmbnail.jpg".
    - Size = 160 x 90 pixels
  - Filename "Video_xxxxxxx_DJI_XXX_mp4_xxxxxxxxxx_xxxxxx_YYYYMDDxxxxxx_photo_preview.jpg".
    - Size = 960 x 540 pixels

The lowercase "x" denotes a sequence of numbers while a capital "X" correlates with the number assigned to the related full-sized photo/video file. The length of the assignment number depends on where the sequence is at.

*c)    Flight Logs*

- Flight records discovered in the root directory of the internal memory when the viewable partition was investigated using Windows Explorer. File details were:
  - Filename "DJIFlightRecord_YYYY-MM-DD_[HH-MM-SS].txt"
  - The created timestamp was accurate.
  - Similar files were found on the RC in "DJI RC\Android\data\dji.go.v5\files\FlightRecord".
  - The log file was successfully parsed in Flight Reader. Figure 18 shows a screen snippet of the application output:

| Date/Time | Path | Event |
|---|---|---|
| 2023-01-28 14:16:08 | /Android/data/dji.go.v5/cache/master | File Created |
| 2023-01-28 14:16:08 | /Android/data/dji.go.v5/cache/master/.nomedia | File Created |
| 2023-01-28 14:16:08 | /Podcasts | File Created |
| 2023-01-28 14:16:08 | /Ringtones | File Created |
| 2023-01-28 14:16:08 | /Alarms | File Created |
| 2023-01-28 14:16:08 | /Notifications | File Created |
| 2023-01-28 14:16:08 | /Pictures | File Created |
| 2023-01-28 14:16:08 | /Movies | File Created |
| 2023-01-28 14:16:08 | /Download | File Created |
| 2023-01-28 14:16:08 | /DCIM | File Created |

Fig. 15. Screenshot taken of Autopsy timeline showing several files and directories created on the SD card at the same time.

| Date/Time | Description | Event Type |
|---|---|---|
| 2023-01-28 18:43:26 | /Android/data/dji.go.v5/files/Editor | File Modified |
| 2023-01-28 18:43:26 | /Android/data/dji.go.v5/files/Editor/music_sound_wave | File Modified |
| 2023-01-28 18:43:26 | /Android/data/dji.go.v5/files/Editor/music_sound_wave | File Accessed |
| 2023-01-28 18:43:26 | /Android/data/dji.go.v5/files/Editor | File Created |
| 2023-01-28 18:43:26 | /Android/data/dji.go.v5/files/Editor/music_sound_wave | File Created |
| 2023-01-28 18:43:18 | /Android/data/dji.go.v5/files | File Modified |

Fig. 16. Screenshot take of Autopsy timeline showing the date and time the RC's power-on sound file was last accessed.

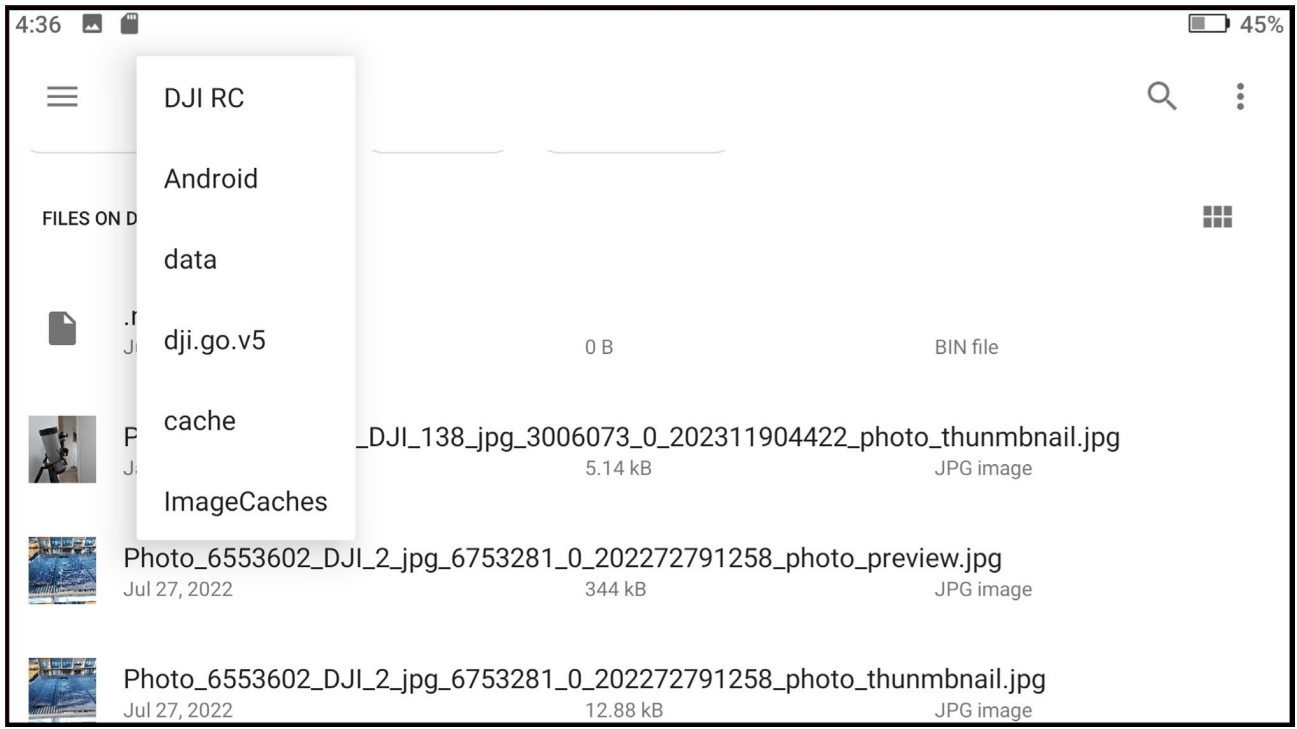

Fig. 17. Thumbnail images discovered on the RC's internal memory via the RC files utility.

Fig 18. Screenshot taken of Flight Reader parsing the recovered flight log.

- o This log displayed the error messages that were shown on the RC screen during the flight.
- o The flight logs contained hundreds of parameters, including flight start and finish time, drone make and model, drone and battery serial numbers, error messages, warnings, GPS coordinates, and altitude.
- o The flight path was also visualised in Google Earth Pro by selecting the KML file download option in Flight Reader. The flight path presented by both applications appeared to be accurate, however, as of the time of writing, neither application appeared to offer start-to-finish visual playbacks of flight paths.

d) *Other*

- The flight for Scenario B could be viewed directly on the RC (via the DJI Fly App) by selecting "Profile", then "more" and then tapping on the flight. The application incorrectly reported the drone model as a "Phantom 4"; however, the flight path and time were accurate.
- The RC-based DJI Fly app listed the email address of the logged in DJI account in "Profile" settings.
- The email associated with the DJI user account was discovered in several text files stored in the volume parent folder "SyncResult". Figure 19 shows a snippet of this text. The timestamps in the filename and within the file indicate when the RC was last synced with DJI's online servers.

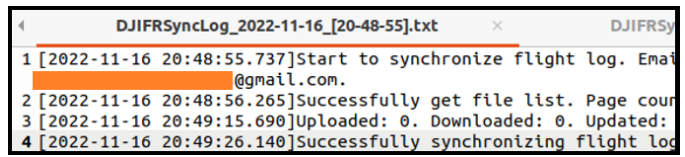

Fig. 19. An email address belonging to the DJI user account signed into the RC was discovered in a log file on the RC's internal memory. Note that the email address has been partially redacted.

- A partial .DAT file was extracted from the RC using DJI Assistant 2 over a USB connection. Further analysis of this file in Notepad++ indicated it was a proprietary encoded binary file. The RC's model number and what appeared to be a serial number were identified in the beginning of the file. See Figure 20. The longer string (a mix of letters and numbers) did not match any serial number identified during the study.

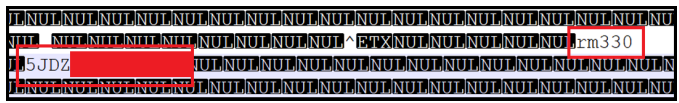

Fig. 20. Model number and unknown string discovered in the .DAT file in plain text. Note that the identified string has been partially redacted.

C. Scenario C

In this scenario, both the drone and RC were recovered. Both devices contained an SD card, and the flight was normal.

1) SD Card – Drone:

   a) *Media*
   - A full-sized MP4 video was retrieved from folder "DCIM\100MEDIA\"
   - Two thumbnail images (a .THM and a .SCR file) of the first frame of the above video were retrieved.

   b) *Flight Logs*
   Nil found.

   c) *Other*
   A subtitle file (.SRT) of the associated video was retrieved in full.

2) Internal Memory Analysis – Drone:

   a) *Media*
   Nil found.

   b) *Flight Logs*
   Nil found.

   c) *Other*
   A .DAT binary file containing the drone's model number and a possible serial number was discovered using DJI Assistant 2 over USB. See Figure 21. The longer string did not match any serial number identified during the study.

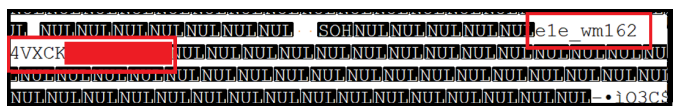

Fig. 21. Model number and possible serial number in binary file extracted from the drone's internal memory. Note that the unidentified string has been partially redacted.

3) SD Card – RC:

   a) *Media*
   Like scenario B, a cached video of the video recorded for scenario C was found on the SD card. The related video of which was discovered on the SD card used in the drone.

   b) *Flight Logs*
   Nil found.

   c) *Other*
   A subtitle file (.SRT) of the associated video was retrieved in full.

4) Internal Memory Analysis – RC:

   a) *Media*
   Thumbnail images of recorded video and image from Scenario C was discovered on the RC in "DJI RC\Android\data\dji.go./v5\cache\ImageCaches"

   b) *Flight Logs*
   Like Scenario B, flight logs were discovered in txt filetype.

   c) *Other*
   - Similar to Scenario B, the flight could be viewed directly on the RC (in the DJI Fly App) by selecting "Profile", then "more" and then tapping on the flight. The application incorrectly reported the drone model as a "Phantom 4"; however, the flight path and time were accurate.
   - Using manual extraction methods, a DJI user account and email address were discovered signed into the DJI Fly app on the RC.

D. Scenario D

For this scenario, both the drone and RC were recovered, however neither contained an SD card. To simulate a collision, the drone was carefully turned over by hand just prior to being powered down. USB connections were not available for this scenario, only Wi-Fi and Bluetooth. Additionally, the DJI user account was logged out of the RC during the flight for this scenario.

1) Internal Memory Analysis – Drone:

   a) *Media*
   - A full-sized image and video taken for this scenario was successfully retrieved from the drone using "Quick Transfer Mode" on the iOS version of the DJI Fly application. The file names of these files were:
     - "dji_fly_YYYYMMDD_xxxxxx_XXX___photo.jpg"

- "dji_fly_YYYYMMDD_xxxxxx_XXX___video.mp4".

*b) Flight Log*

Nil found.

*c) Other*

Nil found.

2) *Internal Memory Analysis – RC:*

*a) Media*
- The "Flyshare" option on the iOS version of the DJI Fly was connected to the DJI Fly app on the RC. No media was presented to the mobile DJI Fly app using this option.
- Using the native Android filesystem utility on the RC, thumbnail images associated with a recorded video and image taken for Scenario D were discovered within the folder "DJI RC\Android\data\dji.go.\v5\cache\ImageCaches".

*b) Flight Logs*
- Using the native Android filesystem utility, a .DAT based flight log for this scenario was discovered in "DJI RC\Android\data\dji.go.b5\files\FlightRecord".
This file could not be transferred via Bluetooth in its native form. The filesystem utility displayed an error on the RC screen "no apps can perform this action". Despite this, the the file could be renamed to a .TXT filetype using the filesystem utility and then transferred to the forensic workstation via Bluetooth. The file was reverted to the original .DAT filetype and converted using DatCon. A .CSV and .txt file were generated by DatCon. Attempts were made to load these files into the Flight Reader application, however neither were successfully parsed. The application did not provide a reason why it could not parse the files. The .DAT file was uploaded to the Airdata UAV website, which it parsed; however, the output did not contain an entry relating to the simulated collision. Figure 22 is a screenshot of some of the information parsed by the Airdata UAV site.
- A .TXT flight log for the flight associated with Scenario D was also discovered in folder "DJI RC\Android\data\dji.go.b5\files\FlightRecord"
using the native Android filesystem utility on the RC. This was able to able to be transferred to the forensic workstation via Bluetooth and parsed by the Flight Reader application. The error associated with the simulated collision during Scenario D was captured by Flight Reader (see Figure 23).
- The same .TXT log was able to be parsed by the Airdata UAV site. Despite being significantly smaller than the .DAT file, it displayed more details, including the same warning message that Flight Reader parsed.

*c) Other*
- The .DAT flight log file contains the "Flight Controller" serial in plain text. This was visually confirmed with serial listed in the "About" page of the RC version of the DJI Fly application.
- The flight history for this scenario was not available for viewing on the iOS version of the DJI Fly app. Note that the auto syncing flight logs option was not enabled on the RC.
- The flight for Scenario D appeared on the RC under "profile" and "more" options when the DJI user account was logged back in.
- The email address of the previously logged in user account was shown in full as an auto-complete prompt on the RC-based DJI Fly app when the first few letters of the email was entered into the username field.

E. *Anti-forensic Methods*

Potential anti-forensic methods to avoid or limit the generation, storage, and transmission of digital artefacts of interest were identified with the Mini 3 Pro and DJI RC. These included:

*1) Purging cached flights from the RC via the DJI Fly application:* It was found that cached flights could be deleted by several methods:
- Remove individual flights from the device by selecting "Profile", then "More", followed by swiping left on the flight to remove, and pressing the red icon depicted with a trash can that appears.
- To bulk clear the flight log cache, select "Profile", then "More", then "Storage". In the box titled "RC Internal Storage", expand the "Clear Cache" menu, and select "Clear Aircraft Flight Record Cache". If applicable, this method can also be used for the SD Card.

*2) Preventing the RC and iPhone from syncing flight records with DJI's online servers:* Two methods were discovered for this:
- Opening the DJI Fly app on the RC, selecting "Profile", then "Settings", then "Sync Flight Data", and disabling the setting "Auto-sync Flight Records".
- Similar steps to the RC, but performed on the iOS version of the DJI Fly app.

*3) Preventing or limiting PII being sent to DJI:* it appeared this could be achieved with the RC's software in several different ways:
- In the DJI Fly app, select "Profile", then "More" then "Notification and Privacy". From there is an option to enable/disable network data transmission labelled "Local Data Mode". Enabling this generated an on-screen prompt that informs the user that all network data transmission would be disabled. These claims were not examined or verified during the experiment.

- In the same menu area as above was a section titled "Privacy", with four different slide switches labelled "mobile Device GPS info", DJI Device GPS Info", "DJI Device Hardware Info", and "Approximate Location Info". It appeared that a user could only disable three of these four 'privacy' options. The exception was the option labelled "Approximate Location Info". Attempts to disable this feature resulted in an immediate on-screen warning prompt with the message "Authorization required for the app to run normally and cannot be disabled". The DJI Security Whitepaper (v2) confirms that this option cannot be disabled [27]. Additionally, it was found that disabling the option "DJI Device Hardware Info" resulted in regular on-screen prompt messages when the drone was powered on and which required the user to acknowledge to remove them. It was found that these warning prompts stopped while the drone was in flight but returned as soon as the drone landed.
- Syncing locally stored data with internet-connected services via the RC OS. It appeared that the controllers' Wi-Fi system could be disabled at the OS level by turning off the Wi-Fi option presented by the OS. This option was located by simply swiping down from the top of the RC screen. This option was not examined thoroughly, nor verified during the experiment.
- Not logging into the DJI Fly app on the RC. It was found that the user account that was previously logged in could be logged out of the application. The drone could still be flown without a user being logged into the DJI Fly application.

*4) Formatting the SD card used in the RC/Drone:* Options to format an SD card mounted on the RC, and the RC internal memory, were discovered on the DJI Fly app and the Android OS filesystem utility. Due to time constraints, the effectiveness of these formatting options was not assessed. It was found however during the study that formatting an SD card previously used in the drone or RC using the SD Association's tool "SD Card Formatter" did not effectively wipe the data from the SD card. This indicates that formatting an SD card is not a reliable method for data sanitization.

*5) Unbinding the device with a DJI account:* It was discovered that a DJI account binded to the drone could be unlinked via the RC- and iOS-based versions of the DJI Fly application by selecting "Profile", then "More", then "Device Management", then selecting "Remove Device from Account" in the "Account and Device" menu. This option was outlined in the official DJI Mini 3 Pro user manual [19]. The DJI Fly application informed the researcher that unbinding the DJI account with the drone would limit the ability to check the device usage. This option was not attempted or verified during the research.

*F. Summary*

Several key digital artefacts were retrieved from both the Mini 3 Pro and the RC. Table 2 provides a summary of the digital artefact findings. Additionally, several potential anti-forensic methods were identified; however, due to time constraints, the efficacy of these methods was not thoroughly assessed during the study.

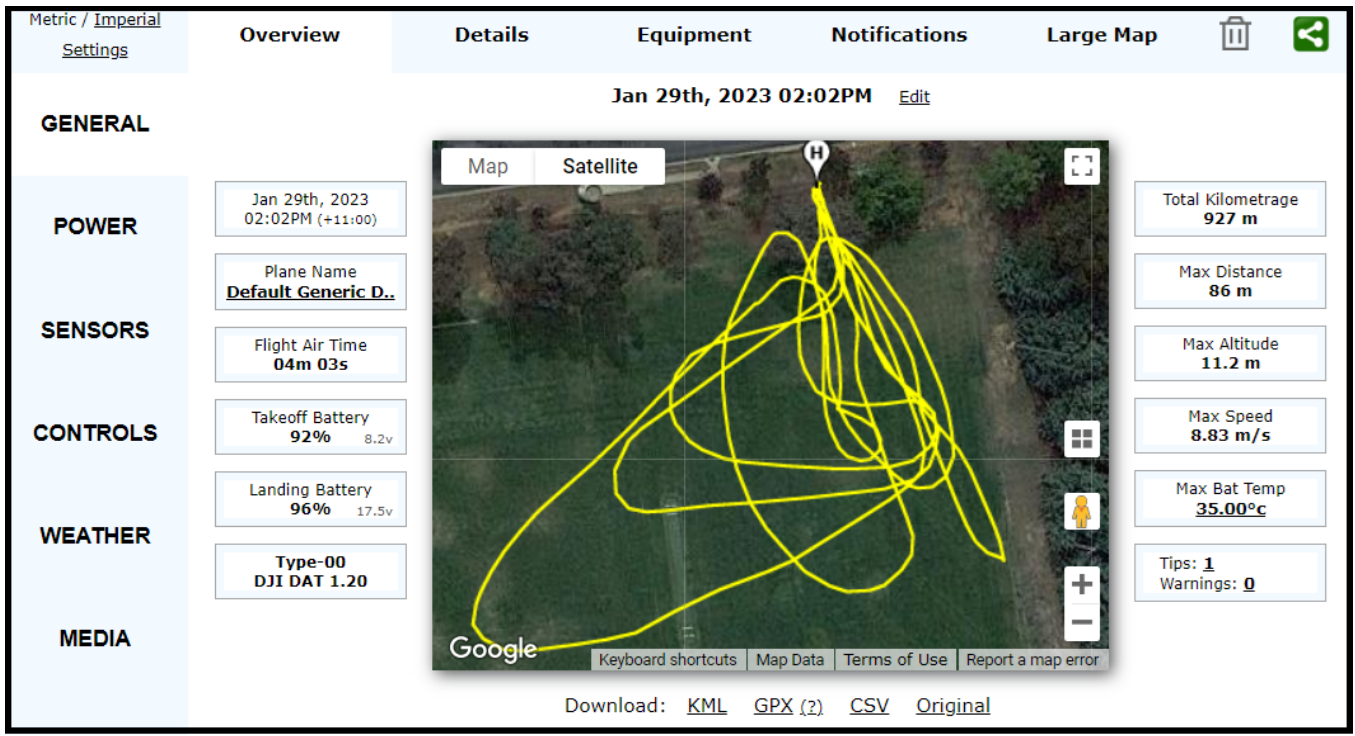

Fig. 22. Screenshot of flight log information decrypted by the Airdata UAV website.

> Excessive aircraft rollover time. Restart aircraft and ensure it is level before taking off. (Code: 16100064).; AIMU attitude restricted.

Fig. 23. Screenshot of flight log information decrypted by Flight Reader.

TABLE 2
A SUMMARY OF THE DIGITAL ARTEFACT TYPES DISCOVERED

| Artefact type: | | Media | | Flight log | | Other | |
| --- | --- | --- | --- | --- | --- | --- | --- |
| Scenario: | Device: | SD | Internal | SD | Internal | SD | Internal |
| A | Drone | Y | N | N | N | G, D | D, E |
| B | RC | Y | Y | N | Y | N | G, P, D |
| C | Drone | Y | N | N | N | G, D | D, E |
|   | RC | Y | Y | N | Y | N | G, P, D |
| D | Drone | -- | Y | -- | N | -- | G, D |
|   | RC | -- | Y | -- | Y | -- | D, E |

Y = Artefact discovered, N = No artefacts discovered, G = Artefacts containing geolocation found, P = Artefacts containing PII found,
D = Artefacts with device identifiable information found, E = Encrypted/encoded artefacts found that could not be decrypted/decoded.

## V. DISCUSSION

Several digital artefacts of potential interest to forensic investigators and researchers were discovered on the Mini 3 Pro and DJI RC. These artefacts included media, flight logs, and other PII, including an email address of the associated drone operator. These results are similar to that of case studies of older DJI drone models by [4], [5], [10], [15]

Acquisition of key digital artefacts from these devices was straightforward, with the majority able to be acquired from simple manual extraction methods, while other artefacts, such as flight logs and previously deleted files, required more effort. The was partially attributed to the absence of password protection on these devices and is likely to assist investigators conducting DRF examination on said devices in the future.

Flight logs located and retrieved from the internal storage of the RC were successfully decrypted by DatCon and Flight Reader. Flight logs decrypted by DatCon were submitted to the Airdata UAV website to present the logs in a visual form, while Flight Reader was capable of decrypting and presenting logs directly to visual form. Both methods were able to present the relevant flight history information in an intuitive way; however, it was discovered that an error log entry relating to the simulated collision in Scenario D was missing from the information presented by Airdata UAV. This is a significant finding as the absence of such information could be crucial in determining the circumstances behind a drone related event. This finding highlights a potential limitation in the capabilities of Airdata UAV's flight log parsing service which investigators should be aware of.

The reviewed literature assessed the capabilities of online services such as Airdata and Google Earth for parsing flight logs and presenting them in a visual form. Uploading flight logs to third parties could raise privacy and legal concerns for investigators. In contrast, this study also evaluated the capabilities of Flight Reader for local flight log analysis. It was found that this application was capable of parsing and presenting DJI flight logs, however, an Internet connection was required to obtain maps and other data. The application developer confirmed that Flight Reader requires an Internet connection to obtain map data, decryption keys, and verify the application licence [28]. Due to this fact, the exchange of such data could also raise privacy concerns for investigators.

Previously deleted subtitle telemetry and image files were recovered from unallocated portions of the SD cards. It is believed this was due to a limitation of the card formatting utilities used in the study to properly sanitise data on the SD cards. This finding was identified early, during analysis of Scenario A, and steps were taken to ensure that the SD cards were sufficiently sanitised for subsequent scenarios.

A cached video file discovered on the RC during analysis of Scenario B was only 11% of the full-sized associated video saved to the drone, however the resolution of the video was sufficient enough to make out details. This finding could be useful to investigators in situations where videos have been recorded, but the associated full-sized video from the drone is unable to be recovered.

It was surprising to see DJI's Wi-Fi-based "Quick Transfer" feature functioned between the drone and RC even though Wi-Fi appeared to be turned off on the RC's Android settings. The reason for this is currently unknown, though it is suspected that the DJI Fly app enables Wi-Fi, Ocusync, and/or Bluetooth to facilitate this media transfer system.

Although it appeared there was an option to disable "Approximate Location" in the RC DJI Fly application settings, it did not appear that it could be disabled. This finding was an unusual juxtaposition considering that the application developers could remove the faux option altogether.

The nature of the strings found in the .DAT files extracted via the DJI Assistant 2 application was unknown to the author. It is proposed that these strings may be hidden serial numbers unique to the respective devices. A comparison with other DJI Mini 3 Pro's and RC's might support or disprove this theory.

Of the two devices, it was discovered that the RC contained all of the PII related digital artefacts. This finding supports the view by [4] that attribution is more likely to be achieved with data acquired from an RC rather than a drone.

As the marketed weight of the DJI Mini 3 Pro is 249 grams [21], this study partially addresses claims by [10] of a gap in the literature pertaining to DRF case studies of drone models weighing less than 250 grams.

The system logs were not able to be decoded, which mirrored the experience of [10]. It would be useful to be able for investigators to be able to decode these.

## VI. FUTURE WORK

Several promising areas for future research were identified out of this case study. The first is further investigation into DJI's DroneID technology. The presence and specifics of the technology on the devices examined in this study are unknown, and as such, DRF case studies that focus on this technology could advance the literature. Inability to decode DJI system logs was a limitation identified during the study. Such logs could provide investigators with valuable insight into the drone's usage and assist with attribution. Therefore, research into methods for decoding these logs could be valuable to digital forensic investigators.

Other potential areas to research include submitting the DJI Mini 3 Pro and DJI RC to vulnerability testing to get root access, performing a chip-on assessment on the devices, assessing the weight carrying capabilities of the drone and investigating the potential presence of digital artefacts relating to a history of payload transport.

## VII. CONCLUSION

In this case study, the DJI Mini 3 Pro and DJI RC were examined for presence of digital artefacts that could be of key interest to investigators. Additionally, potential digital anti-forensic methods for both devices were explored. The case study was divided into four potential DRF scenarios that a digital forensic investigator may be faced with. The study found digital artefacts, such as media, flight logs, and PII, on both devices that could potentially reveal the drone's usage history and identify the operator/registered owner. Of these, flight logs and PII were discovered on the RC, while full-sized media was discovered on the drone. Of the two devices, it is apparent that the RC is the primary device for acquiring drone usage and PII related data. Key limitations encountered during the study included not being able to be decode system logs and not having full access to the internal memory of both devices for complete physical data extraction and analysis. Despite these limitations, the findings in this study could still potentially assist investigators and researchers with future DRF examinations of these and similar devices. In addition, this study partially addresses the previously discussed gap in literature pertaining to DF of drones weighing less than 250g.

## APPENDIX

Table 3 contains a list of abbreviations used in the paper.

TABLE 3
ABBREVIATIONS USED IN THE ARTICLE

| | |
|---|---|
| ACPO | Association of Chief Police Officers |
| AF | Anti-forensics |
| DF | Digital Forensics |
| DRF | Drone Forensics |
| DJI | Da-Jiang Innovations |
| DNG | Digital Negative |
| FAA | Federal Aviation Administration |
| EXIF | Exchangeable Image File Format |
| JPEG | Joint Photographic Experts Group |
| NIST | National Standards and Technology |
| OS | Operating System |
| RC | Remote Controller |
| Rf | Radiofrequency |
| SD | Secure Digital |
| SDR | Software Defined Radio |
| UART | Universal Asynchronous Receiver Transmitter |
| UAV | Unmanned Aerial Vehicles |
| USB | Universal Serial Bus |
| VM | Virtual machine |